# Discretization of Time Series Data

*Elena S. Dimitrova*[*], *John J. McGee, Reinhard C. Laubenbacher*

*Virginia Bioinformatics Institute at Virginia Tech, Bioinformatics Facility, Washington St. (0477), Blacksburg, VA 24061, USA*

## ABSTRACT

Data discretization, also known as binning, is a frequently used technique in computer science, statistics, and their applications to biological data analysis. We present a new method for the discretization of real-valued data into a finite number of discrete values. Novel aspects of the method are the incorporation of an information-theoretic criterion and a criterion to determine the optimal number of values. While the method can be used for data clustering, the motivation for its development is the need for a discretization algorithm for several multivariate time series of heterogeneous data, such as transcript, protein, and metabolite concentration measurements. As several modeling methods for biochemical networks employ discrete variable states, the method needs to preserve correlations between variables as well as the dynamic features of the time series. A C++ implementation of the algorithm is available from the authors at http://polymath.vbi.vt.edu/discretization.

## 1 INTRODUCTION

Discretization of real data into a typically small number of finite values is often required by machine learning algorithms (Dougherty *et al.*, 1995), Bayesian network applications (Friedman and Goldszmidt, 1996), and any modeling algorithm using discrete-state models. Binary discretizations are the simplest way of discretizing data, used, for instance, for the construction of Boolean network models for gene regulatory networks (Kauffman, 1969; Albert and Othmer, 2003). The expression data are discretized into only two qualitative states as either present or absent. An obvious drawback of binary discretization is that labeling the real-valued data according to a present/absent scheme generally causes the loss of a large amount of information. Discrete models and modeling techniques allowing multiple states have been developed and studied in, *e.g.*, (Laubenbacher and Stigler, 2004; Thieffry and Thomas, 1998).

But experimental data are typically continuous, or, at least, represented by computer floating point numbers. For the case of small samples of biological data, many statistical methods for discretization are not applicable due to the insufficient amount of the data. Some other existing discreti-zation techniques assume that the number of discrete classes to be obtained is given, *e.g.*, (Friedman *et al.*, 2000). While this number is extremely important, it is not clear how to properly select it in many cases.

In this paper we introduce a new method for the discretization of experimental data into a finite number of states. While of interest for other purposes, this method is designed specifically for the discretization of multivariate time series, such as those used for the construction of discrete models of biochemical networks built from time series of experimental data. We employ a graph-theoretic clustering method to perform the discretization and an information-theoretic technique to minimize loss of information content. One of the most useful features of our method is the determination of an optimal number of discrete states that is most appropriate for the data. Our C++ program takes as input one or more vectors of real data and discretizes their entries into a number of states that best fits the data. Our main objective was to construct a method that preserves correlations between variables as well as information about network dynamics inherent in the time series. We have validated the method in two ways: by using published DNA microarray data to test for preservation of correlations and by comparing the dynamics of a discrete and continuous model constructed using the modeling method in (Laubenbacher and Stigler, 2004).

## 2 DISCRETIZATION PROBLEM

In order to place our method in a general context we first give a definition of discretization (Hartemink, 2001).

A *discretization* of a real-valued vector $v = (v_1,...,v_N)$ is an integer-valued vector $d = (d_1,...,d_N)$ with the following properties:

(1) Each element of $d$ is in the set $\{0, 1,..., D - 1\}$ for some (usually small) positive integer $D$, called the *degree* of the discretization.

(2) For all $1 \le i, j \le N$, we have $d_i \le d_j$ if $v_i \le v_j$.

Spanning discretizations of degree $D$ are a special case of the discretizations we consider in this paper. They are defined in (Hartemink, 2001) as discretizations that satisfy the additional property that the smallest element of $d$ is equal to 0 and that the largest element of $d$ is equal to $D - 1$. Both methods, however, assume extra knowledge about the data

---
[*] Corresponding author. Tel.: 540-231-3965; Fax: 540-231-2606

*Email addresses:* edimitro@vbi.vt.edu (Elena Dimitrova), jmcgee@vbi.vt.edu (John McGee), reinhard@vbi.vt.edu (Reinhard Laubenbacher).

source, which may not always be available. For example, the sample size may be insufficient to estimate distributions. For time series of transcript data the number of time points is typically much smaller than the number of genes considered, so that statistical approaches to discretization become problematic. Also, in these cases it is rarely known what the appropriate discretization thresholds for each gene might be. Another common discretization technique is based on *clustering* (Jain and Dubes, 1988). One of the most common clustering algorithms is the *k-means clustering* developed by MacQueen (1967). The goal of the *k*-means algorithm is to minimize dissimilarity in the elements within each cluster while maximizing this value between elements in different clusters. The algorithm takes as input a set of points $S$ to be clustered and a fixed integer $k$. It partitions $S$ into $k$ subsets by choosing a set of $k$ cluster *centroids*. The choice of centroids determines the structure of the partition since each point in $S$ is assigned to the nearest centroid. Then for each cluster the centroids are re-computed based on which elements are contained in the cluster. These steps are repeated until convergence is achieved. Many applications of the *k*-means clustering such as the *MultiExperiment Viewer* (Saeed *et al.*, 2003) start by taking any random partition into $k$ clusters and computing their centroids. As a consequence, a different clustering of $S$ may be obtained every time the algorithm is run. Another inconvenience is that the number $k$ of clusters to be formed has to be specified in advance.

Another method is *single-link clustering* (SLC) with the Euclidean distance function on vectors of real data to produce a spanning discretization. SLC is a divisive (top-down) hierarchical clustering that defines the distance between two clusters as the minimal distance of any two objects belonging to different clusters (Jain and Dubes, 1988). In the context of discretization, these objects will be the real-valued entries of the vector to be discretized, and the distance function that measures the distance between two vector entries $v$ and $w$ will be the one-dimensional Euclidean distance $|v - w|$. Top-down clustering algorithms start from the entire data set and iteratively split it until either the degree of similarity reaches a certain threshold or every group consists of one object only. For the purpose of data analysis, it is impractical to let the clustering algorithm produce clusters containing only one real value. The iteration at which the algorithm is terminated is crucial since it determines the degree of the discretization, and one of the most important features of our discretization method is a definition of the termination criteria.

SLC with the Euclidean distance function satisfies one of our major requirements: very little starting information is needed – only distances between points. It may result, however, in a discretization where most of the points are clustered into a single partition if they happen to be relatively close to one another. This negatively affects the information content of the *discrete vector* (to be discussed later in the paper). Another problem with SLC is that its direct implementation takes $D$, the desired number of discrete states, as an input. However, we would like to choose $D$ as small as possible, without losing correlation and dynamic information, so that an essentially arbitrary choice is unsatisfactory. These two issues were addressed by modifying the SCL algorithm: our method begins by discretizing a vector in the same way as SLC but instead of providing $D$ as part of the input, the algorithm contains termination criteria which determine the appropriate number $D$. After that each discrete state is checked for information content and if it is determined that this content can be considerably increased by further discretization, then the state is separated into two states in a way that may not be consistent with SLC. The details of these procedures are given next.

## 3 METHOD

The method assumes that the data to be discretized consist of one or several vectors of real-valued entries. It is appropriate for applications when there is no knowledge about distribution, range, or discretization thresholds of the data and arranges the data points into clusters only according to their relative distance with respect to each other and the resulting information content. The algorithm employs graph theory as a tool to produce a clustering of the data and provides a termination criterion.

### 3.1 Discretization of one vector

Even if more than one vector is to be discretized, the algorithm discretizes each vector independently and for some applications this may be sufficient. The example of such a vector to keep in mind is a time series of expression values for a single gene. If the vector contains $m$ distinct entries, a complete weighted graph on $m$ vertices is constructed, where a vertex represents an entry and an edge weight is the Euclidean distance between its endpoints. The discretization process starts by deleting the edge(s) of highest weight until the graph gets disconnected. If there is more than one edge labeled with the current highest weight, then all of the edges with this weight are deleted. The order in which the edges are removed leads to components, in which the distance between any two vertices is smaller than the distance between any two components, a requirement of SLC. We define the distance between two components $G$ and $H$ to be $\min\{|g - h| \mid g \in G, h \in H\}$. The output of the algorithm is a discretization of the vector, in which each cluster corresponds to a discrete state and the vector entries that belong to one component are discretized into the same state.

### 3.2 Example

Suppose that vector $v = (1, 2, 7, 9, 10, 11)$ is to be discretized. The corresponding SLC dendrogram that would be obtained by SLC algorithms such as the Johnson's algorithm (Johnson, 1967) is given in Figure 1.

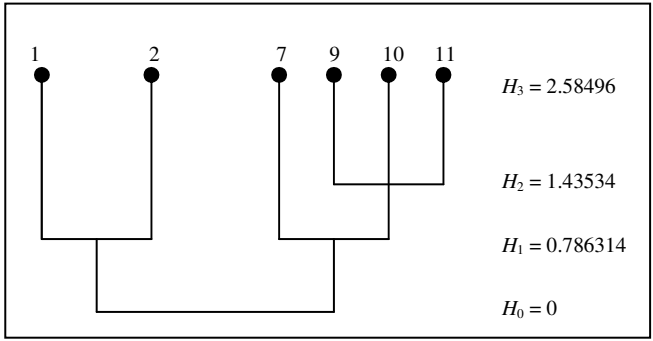

**Fig. 1.** Dendrogram representing the SLC algorithm applied to the data of Example 3.2. The column on the right gives the corresponding Shannon's entropy increasing at each consecutive level.

We start with constructing the complete weighted graph based on $v$ which corresponds to iteration 0 of the dendrogram (Figure 2).

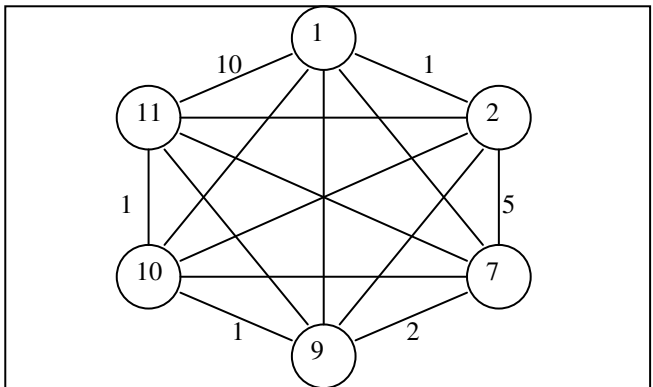

**Fig. 2.** The complete weighted graph constructed from vector entries 1, 2, 7, 9, 10, 11. Only the edge weights of the outer edges are given.

Eight edges with weights 10, 9, 9, 8, 8, 7, 6, and 5, respectively, have to be deleted to disconnect the graph into two components: one containing vertices 1 and 2 and another having vertices 7, 9, 10, and 11; this is the first iteration. Having disconnected the graph, the next task is to determine if the obtained degree of discretization is sufficient; if not, the components need to be further disconnected in a similar manner to obtain a finer discretization. A component is further disconnected if one of the following four conditions is satisfied ("disconnect further" criteria):

(1) The average edge weight of the component is greater than half the average edge weight of the complete graph.

(2) The distance between its smallest and largest vertices is greater than or equal to half this distance in the complete graph. For the complete graph, the distance is the graph's highest weight.

(3) The minimum vertex degree of the component is less than the number of its vertices minus 1. The con-

trary implies that the component is a complete graph by itself, *i.e.* the distance between its minimum and maximum vertices is smaller than the distance between the component and any other component.

(4) Finally, if the above conditions fail, a fourth one is applied: disconnect the component if it leads to a substantial increase in the information content carried by the discretized vector.

The result of applying only the first three criteria is analogous to SLC clustering with the important property that the algorithm chooses the appropriate level to terminate. Applying the fourth condition, the information measure criterion may, however, result in a clustering which is inconsistent with any iteration of the SLC dendrogram. This criterion is discussed next.

### 3.3 Information measure criterion

Discretizing the entries of a real-valued vector into a finite number of states certainly reduces the information carried by the discrete vector in the sense defined by Shannon (1948). In his paper, Shannon developed a measure of how much information is produced by a discrete source. The measure is known as *entropy* or *Shannon's entropy*. Suppose there is a set of $n$ possible events whose probabilities of occurrence are known to be $p_1, p_2, \ldots, p_n$. Shannon proposed a measure of how much choice is involved in the selection of the event or how certain one can be of the outcome, which is given by

$$H = -\sum_{i=1}^{n} p_i \log_2 p_i.$$

The base 2 of the logarithm is chosen so that the resulting units may be called bits.

In our context the Shannon's entropy of a vector discretized into $n$ states is given by

$$H = \sum_{i=0}^{n-1} \frac{w_i}{n} \log_2 \left( \frac{n}{w_i} \right),$$

where $w_i$ is the number of entries discretized into state $i$ (assuming a spanning discretization). An increase in the number of states implies an increase in entropy, with an upper bound of $\log_2 n$. However, we want the number of states to be small. That is why it is important to notice that $H$ increases by a different amount depending on which state is split and the size of the resulting new states. For example, if a state containing the most entries is split into two new states of equal size, $H$ will increase more than if a state of fewer entries is split or if we split the larger state into two states of different sizes.

To see that splitting a given state into two states of equal size results in maximum entropy increase, consider a vector whose entries have been divided into $n$ states, one of which, labeled with 0, contains $w_0$ entries. As a function of $w_0$, the entropy is given by

$$H(w_0) = \frac{w_0}{n} \log_2\left(\frac{n}{w_0}\right) + \sum_{i=1}^{n} \frac{w_i}{n} \log_2\left(\frac{n}{w_i}\right).$$

Suppose that we split state 0 into two states containing $m$ and $w_0 - m$ entries, respectively, where $0 < m < w_0$. This will change only the first term of the right-hand side of the above entropy expression and leave the summation part the same. It is easy to verify that $h(w_0) = \frac{w_0}{n} \log_2\left(\frac{n}{w_0}\right)$ achieves its maximum value over $0 < m < w_0$ at $m = \frac{w_0}{2}$. Therefore, splitting a state into two states of equal size maximizes the entropy increase.

As explained in the previous section, the information measure criterion is applied to a component only after the component has failed the other three conditions. Once this happens, we consider splitting it further only if doing so would provide a very significant increase of the entropy, *i.e.* if the component corresponds to a "large" collection of entries (recurring entries are included since all entries have to be considered when computing the information content of a vector). In our implementation a component gets disconnected further only if it contains at least half the vector entries. Unlike with the other criteria, if a component is to be discretized under the fourth criterion, the corresponding sorted entries are split into two parts: not between the two most distant entries but into two equal parts (or with a difference of one entry in case of an odd number of entries). This is to guarantee a maximum increase of the information measure.

In Example 3.2, the two components that were obtained by removing the edges of heaviest weight both fail the "disconnect further" Conditions $1 - 3$. If the discretization process stopped at this iteration, then the vector $d = (0, 0, 1, 1, 1, 1)$ has Shannon's entropy 0.78631. Having most of the entries of $v$ discretized into the same state, 1, reduces the information content of $d$.

Suppose discretization of $v$ continues according to SLC, *i.e.*, without enforcing the fourth condition of "disconnect further". The next step is to remove the edges of highest weight until a component gets disconnected. This yields the removal of the four edges of weights 4, 3, 2, and 2, respectively, to obtain discretization $d = (0, 0, 1, 2, 2, 2)$. The Shannon's entropy of the new discretization of $v$ is 1.43534. Still half of the entries of $v$ remain at the same discrete level, now 2, which does not allow for a maximal increase in the information content of $d$. If instead discretization proceeded by applying the information criterion to the bigger component, the resulting discretization becomes $d = (0, 0, 1, 1, 2, 2)$ with Shannon's entropy 1.58631, as opposed to the previous entropy of 1.43534.

As illustrated by Example 3.2, the proposed discretization algorithm produces a discretization which is consistent with the definition given above, keeps the number of discrete states small, and maximizes information content over traditional SLC.

### 3.4 Algorithm summary

**Input:** set $S_r = \{v_i \mid i = 1,\dots,m\}$ where each $v_i = (v_{i1},\dots,v_{iN})$ is a real-valued vector of length $N$ to be discretized.

**Output:** set $S_d = \{d_i \mid i = 1,\dots,m\}$ where each $d_i = (d_{i1},\dots,d_{iN})$ is the discretization of $v_i$ for all $i = 1,\dots,m$.

(1) For each $i = 1,\dots,m$, construct a complete weighted graph $G_i$ where each vertex represents a distinct $v_{ij}$ and the weight of each edge is the Euclidean distance between the incident vertices.

(2) Remove the edge(s) of highest weight.

(3) If $G_i$ is disconnected into components $C_{i1}^{G_i},\dots,C_{iM_i}^{G_i}$, go to 4. Else, go to 2.

(4) For each $C_{ik}^{G_i}$, $k = 1,\dots,M_i$, apply "disconnect further" criteria 1–3. If any of the three criteria holds, set $G_i = C_{ik}^{G_i}$ and go to 2. Else, go to 5.

(5) Apply "disconnect further" 4. If criterion 4 is satisfied, go to 6. Else, go to 7.

(6) Sort the vertex values of $C_{ik}^{G_i}$ and split them into two sets: if $|V(C_{ik}^{G_i})|$ is even, split the first $|V(C_{ik}^{G_i})|/2$ sorted vertex values of $C_{ik}^{G_i}$ into one set and the rest – into another. If $|V(C_{ik}^{G_i})|$ is odd, split the first $|V(C_{ik}^{G_i})|/2 +1$ sorted vertex values of $C_{ik}^{G_i}$ into one set and the rest – into another.

(7) Sort the components $C_{ik}^{G_i}$, $k = 1,\dots,M_i$, by the smallest vertex value in each $C_{ik}^{G_i}$ and enumerate them $0, \dots, D_i - 1$, where $D_i$ is the number of components into which $G_i$ got disconnected. For each $j = 1,\dots,N$, $d_{ij}$ is equal to the label of the component in which $v_{ij}$ is a vertex.

### 3.5 Algorithm complexity

Given $M$ variables, with $N$ time points each, we compute $N(N-1)/2$ distances to construct the distance matrix so the complexity of this step is $O(N^2)$. The distance matrix is used to create the edge and vertex sets of the complete distance graph, containing $N(N-1)/2$ edges. This can also be accomplished in $O(N^2)$ time. These edges are then sorted into decreasing order, so that the largest edges are removed first. A standard sorting algorithm, such as merge sort, has complexity $O(N \log N)$ (Knuth, 1998). As each edge is removed, the check for graph disconnection involves testing for the existence of a path between the two vertices of the edge. This test for graph disconnection can be accomplished with a breadth-first search, which has order $O(E+V)$ (Pemmaraju, 2003), with $E$ the number of edges and $V$ the number of vertices in the component. In our case this translates to complexity $O(N^2)$. Edge removal is typically performed for a large percentage of the $N(N–1)/2$ edges, so this step has overall complexity $O(N^4)$. The edge removal step dominates the complexity so that the overall complexity is $O(M\ N^4)$ to discretize all $M$ variables.

While this is the theoretical worst-case performance, because of the heuristics we have added the typical performance is significantly better.

### 3.6 Requirements on the number of states

While for some applications any number of discretization states is acceptable, there are some cases when there are limitations on this number. For example, if the purpose of discretizing the data is to build a model of polynomials over a finite field as in (Laubenbacher and Stigler, 2004), then the number of states must be a power of a prime since every finite field has cardinality $p^n$, where $p$ is prime and $n$ is a positive integer. Our method deals with this problem in the following way.

Suppose that a vector has been discretized into $m$ states in the way described above. The next step is to find the smallest integer $k = p^n$ such that $m \leq k$. This value for $k$ gives the number of states that needs to be obtained. Since the discretization algorithm yielded $m$ clusters, the remaining $k − m$ can be constructed by sorting the entries in each cluster and splitting the one that contains the two most distant entries with respect to Euclidean distance. The splitting should take place between these entries. This is repeated until $k$ clusters are obtained.

This approach has a potential problem. For instance, if a vector got discretized into 14 states and the total number of distinct entries of the vector is 15, then $k = 16$ cannot be reached. In this case the two closest states could be merged together to obtain 13 states. In general it may not be desirable to reduce the number of states because this results in loss of information. We would rather increase the number of states unless it is impossible as in the above example.

### 3.7 Discretization of several vectors

Some applications may require that all vectors in a data set be discretized into the same number of states. For example the approach adopted by Laubenbacher and Stigler (2004) imposes such a requirement on the discretization. The way we deal with this is by first discretizing all vectors separately. Suppose that for $N$ vectors, the discretization method discretized each into $m_1, m_2, \ldots, m_N$ states, respectively. Let $m = \max\{\ m_i \mid i = 1, \ldots, N\ \}$.
Now find the least possible $k = p^n$ such that $m \leq k$. Finally, discretize all variables into $k$ states in the same way that was described for the discretization of a single vector into the required number of states.

## 4 PRESERVATION OF CORRELATIONS

While any discretization inevitably results in information loss, a good multivariate discretization should preserve some important features of the data such as *correlation* between the variables. To demonstrate that our method has this property, we use the temporal map of fluctuations in mRNA expression of a set of genes related to rat central nervous system development presented in (Wen *et al.*, 1998). Wen *et al.* (1998) focus on the cervical spinal cord and the genes included in their study are from families that are believed to be important for spinal cord development. They used an RT-PCR protocol to measure the expression of 112 genes in central nervous system development. The data consist of nine expression measurements for each gene: cervical spinal cord tissue was dissected from animals in embryonic days 11, 13, 15, 18, and 21 and in postnatal days 0, 7, 14, and 90.

We calculated the Spearman rank correlation coefficient between each gene's analog time series and discretized time series according to (Walpole *et al.*, 1998). Then we identified each coefficient value as representing a "significant" or "not significant" correlation based on a critical value of 0.683, which corresponds to a confidence level of 0.025 for a time series of nine points. This means that the probability that the correlation coefficient for a pair of uncorrelated time series of length 9 will be greater than or equal to 0.683 is 0.025 (Walpole *et al.*, 1998). The result is that 71 out of 112 genes were found to be significantly correlated to their discrete version, which is 63.39% of the total number of genes. By augmenting the level of confidence to 0.05, this percentage increases to 76.79%.

We also calculated the Spearman rank correlation coefficient for each pair of genes before and after discretizing the data, considering only the genes that discretized into exactly three states − 56 of them. For the original data given in (Wen *et al.*, 1998), out of the 1540 gene pairs, 234 pairs were identified as significantly correlated with a confidence level of 0.025. Based on the discretized data, 132 pairs were correctly identified as significantly correlated and 1181 pairs were correctly recognized as not significantly correlated. That is, 1313, or more than 85%, of the correlations were correctly classified after discretization.

These results imply that our algorithm can be successfully applied to cases when preserving the relationships between the variables in a system is important.

# 5 APPLICATION TO THE REVERSE ENGINEERING OF AN ARTIFICIAL GENE NETWORK

As mentioned in the introduction, our main motivation was the need to discretize time series in order to construct dynamic models using a finite state set. To demonstrate the effectiveness of our method, we used it to generate appropriate data for the reverse-engineering method developed in (Laubenbacher and Stigler, 2004).

Since data from real gene networks are limited, we chose to test the method on an artificial gene network. We used the A-Biochem software system developed by P. Mendes and his collaborators (Mendes *et al.*, 2003). A-Biochem automatically generates artificial gene networks with particular topological and kinetic properties. These networks are embodied in kinetic models, which are used by the biochemical-network simulator *Gepasi* (Mendes, 1993, 1997) to produce simulated gene expression data. We generated an artificial gene network with five genes and ten total input connections using the Albert-Barabási algorithm (Albert and Barabási, 2000).

*Gepasi* uses a continuous representation of biochemical reactions, based on ordinary differential equations (ODE). With the parameters we specified, *Gepasi* generated an ODE system that represents the network. For example, the synthesis rate of gene $G_1$ is given by

$$\frac{dG_1}{dt} = \frac{0.01\left(\dfrac{G_1(t)}{0.01+G_1(t)}\right)}{0.01+G_3(t)} - G_1(t).$$

Analyzing the dynamics of the ODE system, one finds that it has two stable steady states (of which only the first is biochemically meaningful):
$S_1 = (1.99006, 1.99006, 0.000024814, 0.997525, 1.99994)$ and
$S_2 = (-0.00493694, -0.00493694, -0.0604538, -0.198201, 0.0547545)$.

As Laubenbacher and Stigler (2004) demonstrated, the performance of their algorithm dramatically improves if *knockout* time series for genes are incorporated. For this reason we supplied seven time series of 11 points each: two wild-type time series and five knockout time series, one for each gene. The first wild-type time series is generated by solving the ODE system numerically for $t = 0, 2, 6,..., 20$ with initial conditions $G_i(0) = 1$ for all $i = 1,..., 5$. Figure 3 shows a plot of the numerical solution of the ODE system with these initial conditions. One can simulate a gene knockout in *Gepasi* by setting the corresponding variable and initial condition to zero. The second time series is generated like the first one but this time with $t = 0, 1,..., 10$ and initial conditions $(G_1(0), G_2(0), G_3(0), G_4(0), G_5(0)) = (1, -1, -0.6, -1, 0.5)$.

(We emphasize that we are including the steady state $S_2$ in order to show that the discretized data preserve information about the dynamics of the ODE system.)

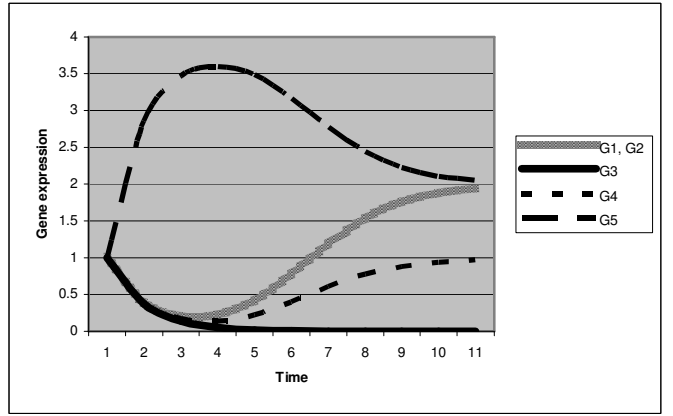

**Fig.3.** Plot of the numerical solution of the ODE system with initial condition $(G_1(0), G_2(0), G_3(0), G_4(0), G_5(0)) = (1, 1, 1, 1, 1)$.

In this case, the time points from each of the seven time series constitute the input vectors. The discretization algorithm chose a state set $X$ of cardinality 5 and, based on the discrete data, the reverse engineering method generated five polynomials describing the discrete model. For example, the polynomial that describes the dynamics of $G_1$ (which is now denoted by x1) is

f1 = -x1*x5^2-2*x2*x5^2+2*x4*x5^2+2*x5^3
    +x1*x2+x2^2-2*x1*x3+x3^2-2*x1*x4+2*x2*x4-
    x3*x4+x4^2-x2*x5+x4*x5-x1-2*x2-x3-2*x5-1.

That is, the discrete model is given by the time-discrete dynamical system

$$f = (f_1, ..., f_5) : X^5 \rightarrow X^5.$$

Now we compare the dynamics of the two models. First, the discretization maps steady state $S_1$ to the fixed point $FP_1 = (4, 4, 1, 4, 2)$ of $f$ and steady state $S_2$ to the fixed point $FP_2 = (0, 1, 1, 1, 0)$. The time series produced by solving the ODE system and converging to $S_1$ is given in the top part of Figure 4. The corresponding discrete points from the time series in the bottom part of Figure 4 form a trajectory that ends at $FP_1$ (Figure 5). The discrete model trajectory can be superimposed over the discretization of the continuous one, illustrating the matching dynamics of the two models. The same can be observed for the second steady-state $S_2$ that is mapped to fixed point $FP_2$.

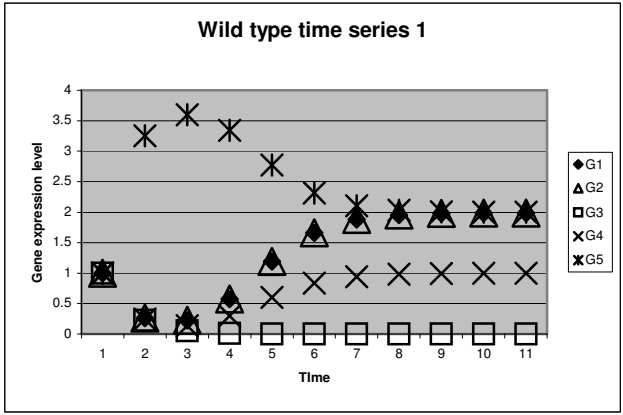

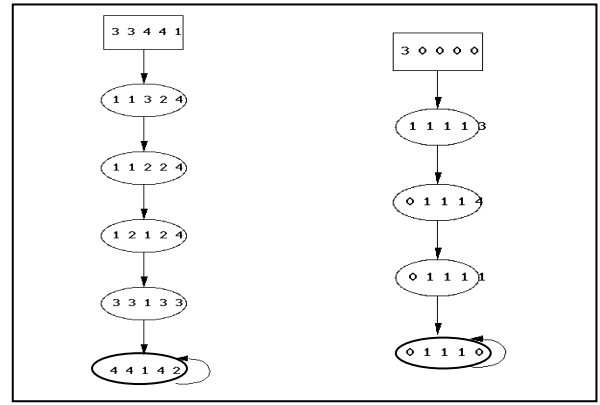

| Time | $G_1$ | $G_2$ | $G_3$ | $G_4$ | $G_5$ |
|------|-------|-------|-------|-------|-------|
| 0    | 3     | 3     | 4     | 4     | 1     |
| 2    | 1     | 1     | 3     | 2     | 4     |
| 4    | 1     | 1     | 2     | 2     | 4     |
| 6    | 1     | 2     | 1     | 2     | 4     |
| 8    | 3     | 3     | 1     | 3     | 3     |
| 10   | 4     | 4     | 1     | 4     | 2     |
| 12   | 4     | 4     | 1     | 4     | 2     |
| 14   | 4     | 4     | 1     | 4     | 2     |
| 16   | 4     | 4     | 1     | 4     | 2     |
| 18   | 4     | 4     | 1     | 4     | 2     |
| 20   | 4     | 4     | 1     | 4     | 2     |

**Fig.4.** Top: wild-type time series generated by solving numerically the ODE system for $t = 0,\ldots,10$ with initial conditions $(G_1(0), G_2(0), G_3(0), G_4(0), G_5(0)) = (1, 1, 1, 1, 1)$; bottom: corresponding discrete point time series.

## 6 DISCUSSION

The discretization method presented here has two novel features. Firstly, it uses Shannon's information criterion to determine clusters and, secondly, it determines the optimal number of clusters for a given data set. In addition to its use as a novel clustering method, it is particularly suitable for the discretization of multivariate time series, since it preserves a large degree of variable correlation and information about dynamic features. It thus provides a valuable tool for any application that requires discretization of continuous data when the number of discrete classes that best fits the data is unknown.

So far our discretization method has only been tested on noiseless data. An important advantage of using discrete states is that a significant portion of the noise is absorbed in the process. The next step is to quantify this portion precisely. Based on preliminary experiments we expect that for data that discretize into a relatively small number of states and that contain a degree of noise common to many biological data, the majority of the noise is absorbed into the discrete states.

## ACKNOWLEDGEMENTS

This work has been partially supported by NIH Grant Nr. 1RO1 GM068947-01. The authors thank A. Jarrah, P. Mendes, and B. Stigler for contributing insights and help with editing, and D. Camacho and W. Sha for helpful discussions.